\documentstyle[10pt,preprint,aps,epsfig]{revtex}
\tightenlines
\newcommand{\bc}{\begin{center}}
\newcommand{\ec}{\end{center}}
\newcommand{\beqn}{\begin{equation}}
\newcommand{\eeqn}{\end{equation}}
\newcommand{\barr}{\begin{eqnarray}}
\newcommand{\earr}{\end{eqnarray}}

\def\etal {{\it et al}. }

\def\tr {\mbox{tr}}

\def\JP #1 #2 #3 {J. Phys.~{\bf#1} (#2) #3}
\def\PL #1 #2 #3 {Phys. Lett.~{\bf#1} (#2) #3}
\def\NP #1 #2 #3 {Nucl. Phys.~{\bf#1} (#2) #3}
\def\NPP #1 #2 #3 {Nucl. Phys.~{\bf B} (Proc.~Suppl.)~{\bf#1} (#2) #3}
\def\ZP #1 #2 #3 {Z.~Phys.~{\bf#1} (#2) #3}
\def\PR #1 #2 #3 {Phys. Rev.~{\bf#1} (#2) #3}
\def\PP #1 #2 #3 {Phys. Rep.~{\bf#1} (#2) #3}
\def\PRL #1 #2 #3 {Phys. Rev.~Lett.~{\bf#1} (#2) #3}
\def\PTP #1 #2 #3 {Prog. Theor.~Phys.~{\bf#1} (#2) #3}
\def\PTPS #1 #2 #3 {Prog. Theor.~Phys.~Suppl.~{\bf#1} (#2) #3}
\def\MPL #1 #2 #3 {Mod. Phys.~Lett.~{\bf#1} (#2) #3}
\def\RMP #1 #2 #3 {Rev. Mod.~Phys.~{\bf#1} (#2) #3}
\def\IJM #1 #2 #3 {Int. J.~Mod.~Phys.~{\bf#1} (#2) #3}
\begin{document}
\draft
\title{Monopole Clustering 
and 
Color Confinement \\
in the Multi-Instanton System}
\author{M.~Fukushima \footnote[3]
{E-mail address:~masa@rcnp.osaka-u.ac.jp}
, H. Suganuma and H. Toki}
\address{\it
Research Center for Nuclear Physics, Osaka University,\\
10-1 Mihogaoka, Ibaraki, Osaka, 567-0047, Japan}
\maketitle
\renewcommand{\thefootnote}{\fnsymbol{footnote}}
\newlength{\minitwocolumn}
\setlength{\minitwocolumn}{7.0cm}
\setlength{\columnsep}{0.8cm}
\renewcommand{\thefigure}{\arabic{figure}} 
\newcommand{\fcaption}[1]{\refstepcounter{figure} Fig.~\thefigure  #1}
\begin{abstract}
\baselineskip 0.5cm
We study color confinement properties of the multi-instanton 
system, which seems to carry an essence of the nonperturbative
QCD vacuum. Here we assume that the multi-instanton system 
is characterized by the infrared suppression of instantons
as $f(\rho)\sim \rho^{-5}$ for large size $\rho$.
We first investigate a monopole-clustering appearing in the
maximally abelian (MA) gauge by considering the correspondence 
between instantons and monopoles. In order to clarify the 
infrared monopole properties, we make the ``block-spin'' 
transformation for monopole currents. The feature of monopole 
trajectories changes drastically with the instanton density. 
At a high instanton density, there appears one very long and 
highly complicated monopole loop covering the entire physical 
vacuum. Such a global network of long-monopole loops resembles 
the lattice QCD result in the MA gauge. Second, we observe 
that the ${\rm SU(2)}$ Wilson loop obeys an area law and the 
static quark potential is approximately proportional to the 
distance $R$ between quark and anti-quark in the multi-instanton 
system using the SU(2) lattice with a total volume of $V=(10 {\rm fm})^4$ 
and a lattice spacing of $a=0.05{\rm fm}$. We extract the string 
tension from the $5 \times 10^{6}$ measurements of Wilson loops.
With an instanton density of $(N/V)=(1/{\rm fm})^{4}$ 
and a average instanton size of $\bar{\rho}=0.4~{\rm fm}$, 
the multi-instanton system provides the string tension of about 
$0.4~{\rm GeV}/{\rm fm}$.
\end{abstract}

\vspace{0.5cm}
\newpage

\baselineskip 0.5cm

\section{Introduction}
Quantum chromodynamics (QCD) has been established as 
the fundamental theory of the strong interaction. In the 
infrared region, there appear various nonperturbative 
phenomena such as color confinement and dynamical 
chiral-symmetry breaking. Since the QCD vacuum is composed 
of gluon fields interacting in a highly complicated way, 
it is hard to understand these phenomena from perturbative 
points of view. On the other hand, a topological aspect 
may provide a useful approach for descriptions of the QCD 
vacuum. Actually, there appear two non-trivial topological 
objects, {\it instantons} and {\it monopoles}, due to the 
non-linearity of QCD.

The instanton configuration discovered in 1975 by Belavin, 
Polyakov, Shvarts and Tyupkin \cite{Polya1} is a classical 
and self-dual solution of the Euclidean field equation in the 
Yang-Mills theory. The appearance of instantons corresponds 
to a homotopy group, $\Pi_{3}({\rm SU}(N_{c}))=
{\rm Z}_{\infty}$ \cite{Rajar}. Instantons are important 
for nonperturbative phenomena related to the ${\rm U_{A}(1)}$ 
anomaly and the large $\eta^{\prime}$ mass \cite{tHoof3}. 
Chiral-symmetry breaking could also be interpreted as the 
instanton effect \cite{Shury4,Diako6,Shury1}. However, until 
now there has been no evidence that instantons have anything 
to do with color confinement in the 4-dimensional gauge theory, 
although Polyakov discovered that instantons cause confinement 
in certain 3-dimensional Georgi-Glashow models \cite{Polya2}. 
With recent computational progress, instanton properties are 
investigated by lattice QCD simulations based on the cooling 
procedures, which are achieved by an artificial reduction of 
the local lattice action. This method allows one to eliminate 
short-range quantum fluctuations of gluon fields and to extract 
only topological excitations like instantons from the 
nonperturbative QCD vacuum 
\cite{Teper,Polik,Campostrini:1990dh,Mich,DeFor2,DeGra2}.
These investigations provide us the average instanton size 
$\bar{\rho} \simeq (0.33-0.4)~{\rm fm}$ and the instanton 
number density $(N/V)\!\simeq \!1~{\rm fm^{-4}}$
\cite{Teper,Polik,Campostrini:1990dh,Mich,DeFor2,DeGra2,Negele:1998ev}. 
Although the existence of instantons with high density 
is agreed, the detailed numbers for $\bar{\rho}$ and
the size distribution depend largely on the cooling 
procedure. These scale parameters suggest that the QCD 
vacuum is the dense matter of instantons and anti-instantons.

In 1981, 't Hooft proposed the abelian gauge, where the 
color-magnetic monopole appears as a relevant degree of freedom 
for the description of color confinement \cite{tHoof1}. In the 
abelian gauge, the ${\rm SU}(N_{c})$ nonabelian gauge theory is 
reduced to the ${\rm U}(1)^{N_{c}-1}$ abelian gauge theory with 
color-magnetic monopoles. The appearance of magnetic monopoles 
corresponds to another homotopy group, $\Pi_{2}({\rm SU}(N_{c})
/{\rm U}(1)^{N_{c}-1})={\rm Z}_{\infty}^{N_{c}-1}$, which is 
different from that of instantons. In the abelian gauge 
the color confinement mechanism can be interpreted as the dual 
Meissner effect due to monopole condensation, which is a dual version 
of Cooper pair condensation in the ordinary superconductivity. 
Such a dual superconductor picture for color confinement was 
proposed by Nambu, 't Hooft and Mandelstam in the middle of 
1970's \cite{Nambu1,tHoof4,Mande1}. By lattice QCD simulations, 
it is observed that large monopole clustering covers the entire 
physical vacuum in the confinement phase, which is identified 
as a signal of monopole condensation being responsible for 
confinement \cite{Kronf1,Kronf2}. Many studies indicate that 
the monopole is a relevant degree of freedom for color 
confinement and chiral-symmetry breaking 
\cite{Brand1,Hioki,Kitah,Miyam2,Wolos,Sugan5,Sasak1,Sasak3}.

Recent studies show remarkable facts that instantons are directly 
related to monopoles \cite{Sugan2,Sugan1,Miyam1,Thurn1,Chern1,Sugan3,Hart,Marku1,Thurn2,Borny,Browe,Polik2,Sugan4,Thurn3,Fukus1,Fukus2,Sasak4,Sasak5,Ilgen1} in the abelian gauge, 
although these topological objects belong to different homotopy 
group. For example, in the Polyakov-like gauge \cite{Sugan2}, 
monopoles appear due to the existence of the hedgehog 
configurations near instanton centers. Such correlations 
indicate that instantons may be important for the promotion of 
long monopole loops. Therefore, we take the multi-instanton system 
instead of the QCD vacuum in order to clarify monopole clustering 
and confinement properties in terms of instantons.

\section{Topological Objects in the QCD Vacuum}
The instanton is a classical and non-trivial solution of the 
Euclidean Yang-Mills theory, whose action is written as $
S = \int d^{4}x G_{\mu\nu}^{a} G_{\mu\nu}^{a}/4g^{2}$.
Here, in order to make the notation and the discussion simpler, 
we take the ${\rm SU(2)}$ case and represent 
$A_{\mu}(x) \equiv ig A_{\mu}^{a}(x)\tau^{a}/2$ and
$G_{\mu\nu}(x) \equiv ig G_{\mu\nu}^{a}(x)\tau^{a}/2$,
which are defined as the anti-hermite variables. Instanton 
and anti-instanton configurations are characterized by the 
(anti-)self-duality condition,
\beqn
G^{a}_{\mu\nu} = \pm \tilde{G}^{a}_{\mu\nu},
\label{eqn:self}
\eeqn 
by using the dual field strength $\tilde{G}_{\mu\nu} \equiv
\frac{1}{2} \varepsilon_{\mu\nu\alpha\beta} G_{\alpha\beta}$.
This condition provides the minimal action $S=\frac{8\pi^{2}}
{g^{2}}\mid{\cal Q} \mid$. Here, the topological charge is 
defined as ${\cal Q} \equiv \int d^{4}x \{ G^{a}_{\mu\nu}
\tilde{G}^{a}_{\mu\nu} \}/32\pi^{2}$
These instanton solutions satisfy the general Yang-Mills field 
equations automatically, $D_{\mu}G_{\mu\nu}\!=\!\pm D_{\mu}
\tilde{G}_{\mu\nu}\!=\!0$. The self-dual solution with 
${\cal Q}\!=\!1$ in the singular gauge \cite{Rajar} is written 
as
\beqn
A^{I}_{\mu}(x;z,\rho,O) =  \frac{\tau^{a}}{2} 
\frac{2iO^{ab}\bar{\eta}^{b\mu\nu}(x-z)_{\nu}\rho^{2}}
{(x-z)^{2}\{ (x-z)^{2}+\rho^{2}\} }.
\label{eqn:ins_solu}
\eeqn
Here, the instanton solutions have several collective modes related 
to the size $\rho$ and the position $z$ of instanton, which are 
described by one and four parameters, respectively. For pure 
${\rm SU(2)}$ gauge theory, the instanton solution can be rotated 
in color space by the color orientation matrix $O$, which is 
characterized by 3 parameters as the Euler angle.
The 't~Hooft symbol $\bar{\eta}^{b\mu\nu}$ is defined as
\barr
\bar{\eta}^{b\mu\nu} = - \bar{\eta}^{b\nu\mu} \equiv \left\{
\begin{array}{ll}
\varepsilon^{b\mu\nu},   &\quad \mu,\nu = 1,2,3 \\
    -\delta^{b\mu},      &\quad \nu     = 4.
\end{array} \right.
\earr
The anti-self-dual solution $A^{\bar{I}}_{\mu}$ is obtained by 
replacing $\bar{\eta}^{b\mu\nu}$ to $\eta^{b\mu\nu}\equiv 
(-1)^{\delta^{\mu 4}+\delta^{\nu 4}}\bar{\eta}^{b\mu\nu}$ in 
Eq.(\ref{eqn:ins_solu}). 

Now, let us discuss the appearance of QCD-monopoles in the 
abelian gauge. We consider the maximally abelian (MA) gauge 
\cite{Kronf1}, which is defined in the Euclidean ${\rm SU}(2)$ 
QCD by minimizing the variable 
\beqn
R_{ch}[A_{\mu}]=  \int d^{4}x
\{ (A^{1}_{\mu}(x))^{2} + (A^{2}_{\mu}(x))^{2} \}
\label{eqn:MAint}
\eeqn
under the gauge transformation $\Omega(\boldmath x)$. 
In the MA gauge, the off-diagonal fields are suppressed 
by the gauge transformation and the full gauge field 
$A_{\mu}=A^{a}_{\mu}\frac{\tau^{a}}{2}$ behaves as the abelian
gauge field $a_{\mu}=A^{3}_{\mu}\frac{\tau^{3}}{2}$ approximately.
In this paper, we adopt the MA gauge, where the 
abelian dominance holds for the nonperturbative QCD phenomena
\cite{Hioki}.
Under the gauge transformation, the gauge field 
transforms as
\barr
A_{\mu}(\boldmath x) \rightarrow A^{\Omega}_{\mu}
(\boldmath x)=\Omega (\boldmath x)(A_{\mu}(\boldmath x) 
+\partial_{\mu})\Omega (\boldmath x)^{\dagger}.
\label{eqn:gautra}
\earr
After the gauge fixing, there only remains the abelian 
gauge symmetry ${\rm U}(1)^{N_{c}-1}\subset{\rm SU}(N_c)$. 
When we meet hedgehog configurations like 
instantons, the multi-valuedness happens to the gauge 
function $\Omega(\boldmath x)$ and the gauge transformation 
of Eq.(\ref{eqn:gautra}) develops a singularity. This 
singularity leads to the monopole current $k_{\mu}$. The 
field strength is defined generally as $G_{\mu\nu}\equiv[
\hat{D}_{\mu},\hat{D}_{\nu}]-[\hat{\partial}_{\mu},
\hat{\partial}_{\nu}]$ \cite{Sugan6}, which returns to the 
standard definition $G_{\mu\nu} = [\hat{D}_{\mu},\hat{D}_{\nu}] 
= \partial_{\mu} A_{\nu}-\partial_{\nu}A_{\mu}+[A_{\mu},
A_{\nu}]$ for the regular case. Under the 
singular gauge transformation $\Omega (x)$, the field 
strength transforms as 
\beqn
G^{\Omega}_{\mu\nu}=\Omega G_{\mu\nu}\Omega^{\dagger}
=\partial_{\mu}A^{\Omega}_{\nu}-\partial_{\nu}A^{\Omega}_{\mu}
+[A^{\Omega}_{\mu},A^{\Omega}_{\nu}]
-\Omega [\partial_{\mu},\partial_{\nu}]\Omega^{\dagger}.
\eeqn
Since the last term is diagonal, abelian field strength is 
naturally obtained as 
\beqn
F_{\mu\nu}=\partial_{\mu}a_{\nu}-\partial_{\nu}a_{\mu}
-\Omega [\partial_{\mu},\partial_{\nu}]\Omega^{\dagger},
\label{eqn:abefiel}
\eeqn
by performing the abelian projection, 
$A_{\mu}^{\Omega}\stackrel{AP}{\longrightarrow} a_{\mu}\equiv 
\tr (A_{\mu}^{\Omega }\tau^{3})\frac{\tau^{3}}{2}$.
The abelian Bianchi identity is broken due to the existence of 
the last term in Eq.(\ref{eqn:abefiel}), and the magnetic current 
$k_{\mu}(x)$ is obtained as 
\barr
k^{\mu}(x)
\equiv \frac{1}{2} \varepsilon_{\mu\nu\alpha\beta} \partial_{\nu}
F_{\alpha\beta}
= - \frac{1}{2} \varepsilon_{\mu\nu\alpha\beta}
\partial_{\nu}\Omega [\partial_{\alpha},\partial_{\beta}]
\Omega^{\dagger}.
\earr
The obvious consequence of the monopole current conservation,
$\partial_{\mu}k^{\mu}=0$, means that monopole currents form
closed loops.

\section{Correlation between Instanton and Monopole for Color Confinement}
Recently, both analytical and lattice studies showed a strong 
correlation between instantons and monopoles in the abelian 
projected theory of QCD 
\cite{Sugan2,Sugan1,Miyam1,Thurn1,Chern1,Sugan3,Hart,Marku1,Thurn2,Borny,Browe,Polik2,Sugan4,Thurn3,Fukus1,Fukus2,Sasak4,Sasak5,Ilgen1}.
Let us briefly review recent studies on the correlation between 
these topological objects in the MA gauge \cite{Hart,Browe}. 
The minimizing condition of $R_{ch}[A_{\mu}]$ in 
Eq.(\ref{eqn:MAint}) satisfies the local condition,
\beqn
(\partial_{\mu} \mp A^{3}_{\mu})A^{\pm}_{\mu}=0,
\hspace{1.0cm} A^{\pm}_{\mu}\equiv A^{1}_{\mu} \pm i A^{2}_{\mu}.
\label{eqn:diff}
\eeqn
It is noted that self-dual solutions like a instanton satisfy 
the stationary condition (\ref{eqn:diff}) automatically. 
However, this stationary condition is not sufficient to realize 
the MA gauge. In the MA gauge the functional $R_{ch}[A_{\mu}]$ must 
be minimized in addition. 
The arbitrary gauge choice of a single instanton 
gauge field leads to a different value of $R_{ch}[A_{\mu}]$, 
although the instanton gauge field satisfies the condition 
(\ref{eqn:diff}) in any gauge \cite{Browe}. 
For instance, in the singular gauge, which has a point 
singularity at the instanton center, the instanton 
configuration gives a finite value $R_{ch}[A_{\mu}^{s}] 
= 4\pi^{2}\rho^{2}$ with $\rho$ being the instanton size. 
On the other hand, in the nonsingular gauge, there appears 
the divergence as $R_{ch}[A_{\mu}^{n}]=2\int d^{4} x 
\frac{\rho^{2}}{(x^{2}+\rho^{2})^{2}}\rightarrow \infty$.
Therefore, it is necessary to consider the minimization of 
the functional $R_{ch}[A_{\mu}]$ for the MA gauge fixing.

Brower \etal consider the single instanton configuration 
in an ``intermediate'' gauge where the gauge field has the 
singularity at the closed loop with a radius $R$ around its 
center \cite{Browe}. This singularity leads a closed monopole 
loop of radius $R$ around the instanton. The parameter $R$ 
should be decided by minimizing the functional $R_{ch}[A_{\mu}]$. 
It is numerically shown that 
the monopole loop with radius $R$ prefers to shrink to a 
point \cite{Browe}. This fact is natural from the following 
consideration. There is no definite direction both in the 
single instanton configuration and in the MA gauge condition 
due to the 4-dimensional rotation invariance. Then, the 
normal vector of the monopole loop cannot be fixed and the 
loop ought to shrink to a point. However, the $R$ dependence of 
$R_{ch}[A_{\mu}]$ is found to be extremely small. Therefore,
some small perturbations as the presence of the finite volume
cause a closed monopole loop with a non-zero radius $R$ due to 
the fragile nature of the monopole point solution, which is 
first reported by Hart and Teper using the lattice 
simulation \cite{Hart}. 
In the QCD vacuum, there are actually many instantons and 
anti-instantons, and here the appearance of monopole loops 
should be influenced by other instantons and anti-instantons. 
Therefore, we would like to investigate the monopole 
loops by using a more realistic multi-instanton system 
defined on the ${\bf R}^{4}$ space.

We first study the monopole clustering as a signal of monopole 
condensation in terms of instantons, which is motivated by these 
strong correlations between instantons and monopoles in the MA 
gauge. Second, we calculate the static quark potential 
in the multi-instanton system by using the Wilson loop in order 
to clarify the relation of instantons with color confinement.

\section{Multi-Instanton Model}
In this section, we would like to model the nonperturbative QCD 
vacuum in terms of instantons. The QCD vacuum possesses the gluon 
condensate \cite{Shifm}, which relates the number of instantons
and anti-instantons \cite{Diako6}. If the total action can be estimated 
as the sum of individual instanton actions, the gluon condensate is 
proportional to the instanton density as $\langle G^{a}_{\mu\nu}
G^{a}_{\mu\nu}\rangle /32\pi^{2}\simeq (N/V)$. The QCD sum rule 
provides the phenomenological value $\langle G^{a}_{\mu\nu}
G^{a}_{\mu\nu}\rangle /32\pi^{2}\simeq (200~{\rm MeV})^{4}$, 
and hence the average instanton density becomes $(N/V) \!\simeq 
\!(1/{\rm fm})^{4}\!$ by the above assumption. The lattice QCD 
simulations suggest that the QCD 
vacuum is saturated with many instantons and anti-instantons. 
Hence, we model the QCD vacuum by the multi-instanton 
ensemble. First of all, we consider the partition function 
${\cal Z}^{1}_{inst}$ of the single instanton as the basic 
ingredient of the multi-instanton theory. Using the collective 
coordinates, $\rho$, $z_{\mu}$, $O$, the single-instanton
partition function \cite{Diako6} is expressed as
\barr
{\cal Z}^{1}_{inst}  = 
\int d^{4}z_{\mu} \int d\rho \int d O f_{0}(\rho) , 
\hspace{0.5cm}
f_{0}(\rho) =\frac{C(N_{c})}{\rho^{5}}[
\frac{8\pi^{2}}{g^{2}(M)}]^{2N_{c}}
(M\rho)^{b}\exp (-\frac{8\pi^{2}}{g^{2}(M)})
\earr
with $b=\frac{11}{3}N_{c}$. Here, $f_{0}(\rho)$ is the 
single-instanton weight function in the one-loop approximation 
\cite{tHoof2,Berna}. This weight function has a scale $M$ 
corresponding to the scale invariance breaking by the trace 
anomaly in QCD. The bare coupling $g^{2}(M)$ is given at this 
scale $M$.  The weight function 
$f_{0}(\rho)$ with $N_{c} \geq 2$ increases with the instanton 
size $\rho$, and the infrared divergence appears in 
${\cal Z}^{1}_{inst}$ within the one-loop approximation.

We consider now the multi-instanton system.
Based on the single-instanton partition function,
the multi-instanton partition function 
is expressed as
\beqn
{\cal Z} \propto
\sum_{N_{+} N_{-}}
\frac{1}{N_{+}!} 
\frac{1}{N_{-}!} 
\prod^{N_{+} + N_{-}}_{n=1}
\int d^{4} z_{n} \int d \rho_{n} \int d O_{n} f_{0}(\rho_{n})
\exp \{- U_{int}(z_{n},\rho_{n},O_{n})\}, 
\label{eqn:part1}
\eeqn
where $U_{int}$ denotes the interaction between instantons 
(anti-instantons) and depends generally on $z_{n}$, $\rho_{n}$, 
$O_{n}$. Here, we follow the standard method, where the 
interaction $U_{int}$ is taken to depend only on the size 
$\rho$ \cite{Diako1,Shury2,Shury3}. Therefore, we consider now
$f(\rho) = f_{0}(\rho)\exp\{-U_{int}\}$ as the instanton size 
distribution. This interaction is found to be repulsive, 
which suppresses the appearance of large size instantons 
\cite{Diako1,Shury2,Shury3}. We regard the positions 
and the color orientations as random variables in our 
calculation.

For the small instanton, the perturbative scheme is valid 
and the interaction $U_{int}$ can be neglected. Hence, the 
instanton size distribution $f(\rho)$ behaves as  
the single instanton weight $f_{0}(\rho)$,
\beqn
f(\rho) \stackrel{\rho \rightarrow 0}{\longrightarrow} 
{\rm const} \cdot \rho^{b-5}.
\label{eqn:dis1}
\eeqn
On the other hand, for the large instanton, the direct 
estimation of $f(\rho)$ is very complicated due to the 
nonperturbative properties. The analytical studies 
and the numerical lattice QCD calculations \cite{Diako2,Shury6} 
suggest a strong suppression of the large size instanton as
\beqn
f(\rho) \stackrel{\rho \rightarrow \infty}{\longrightarrow} 
{\rm const} \cdot \rho^{-\nu},
\label{eqn:dis2}
\eeqn
which is caused by the repulsive force in the infrared 
region. The ordinary instanton liquid model suggests 
$\nu =5$ \cite{Shury6}. 

To connect the two tendencies in Eqs.(\ref{eqn:dis1}) and 
(\ref{eqn:dis2}) smoothly, we take the size distribution as
\begin{equation}
f(\rho )=\frac{1}{(\frac{\rho}{\rho_{1}})^{\nu }+
(\frac{\rho_{2}}{\rho})^{b-5} }
\label{eqn:distri}
\end{equation}
where $\rho_{1}$ denotes the infrared size parameter 
and $\rho_{2}$ the ultraviolet size parameter. These 
two parameters are fixed by the average instanton size 
$\bar{\rho}\equiv \int^{\infty}_{0} d\rho \rho f(\rho) 
= 0.4~{\rm fm}$ \cite{Shury5} and the normalization 
condition $\int^{\infty}_{0} d\rho f(\rho) = 1$.

The above discussion leads to the use of the partition, 
function
\beqn
{\cal Z} \propto
\sum_{N_{+} N_{-}}
\frac{1}{N_{+}!} 
\frac{1}{N_{-}!} 
\prod^{N_{+} + N_{-}}_{n=1}
\int d^{4} z_{n} \int_{0}^{\infty} d \rho_{n} \int d O_{n}
\frac{1}{(\frac{\rho_{n}}{\rho_{1}})^{\nu }+
(\frac{\rho_{2}}{\rho_{n}})^{b-5} }.
\eeqn
As for the gluon field $A_{\mu}$ of the multi-instanton 
system, we take the sum ansatz \cite{Diako3}, 
\begin{equation}
A_{\mu }(x)=\sum_{k} A_{\mu }^{I}(x;z_{k},\rho_{k},O_{k}) 
+\sum_{\bar{k}} A_{\mu }^{\bar{I}}(x;z_{\bar{k}},
\rho_{\bar{k}},O_{\bar{k}}),
\label{eqn:inssum}
\end{equation} 
which is constructed by the instanton and anti-instanton 
solutions in the singular gauge. Using the variational 
treatment with this ansatz, Diakonov and Petrov have 
shown the appearance of the repulsive force between
instanton and anti-instanton in the infrared region 
\cite{Diako3}.

In actual calculations, we generate the ensemble of 
instantons and anti-instantons with random centers 
$z_{k}$ on the 4-dimensional Euclidean continuum space. 
The color orientations $O_{k}$ are taken randomly.
The instanton sizes $\rho_{k}$ are randomly chosen 
following the size distribution $f(\rho)$ in 
Eq.(\ref{eqn:distri}). In the simple sum ansatz, 
the interaction among these pseudoparticles is 
supposed to be included effectively in the instanton 
size distribution $f(\rho)$.
The instanton size distribution for the cases of $\nu=5$
and $3$ are shown in Fig.\ref{fig:Size_Dis}. 

\section{Monopole Clustering in Multi-Instanton System }
Based on the lattice gauge theory, we investigate 
monopole-loop distributions induced by instantons
after the MA gauge fixing \cite{Fukus1,Fukus2}. 
We introduce a lattice on the multi-instanton 
configuration and define the link variable $U_{\mu }(s)
={\rm exp}(iaA_{\mu }(s))$, where $A_{\mu}(s)$ is 
provided on each link from Eq.(\ref{eqn:inssum}). 
In the actual calculation, we use the $32^{4}$ lattice 
with the lattice-spacing of $a=0.125~{\rm fm}$.

First of all, let us discuss the boundary condition
of the multi-instanton configuration. If instantons
and anti-instantons are generated only in the 
finite volume $V=(32\times a)^{4}=(4.0~{\rm fm})^{4}$, 
the strength of the gluon field is insufficient near 
the border of the volume $V$. To avoid this border 
problem, we adopt the periodic boundary condition, which  
ensures adequate contributions also near the border. 
Here, instantons are assumed to appear with the periodic 
position, size and color orientation out of the finite 
volume $V$. A schematic view of this periodic condition 
of an instanton ensemble sliced into 2-dimensional plane 
is shown in Fig.\ref{fig:Int_M_B_2D}. This means the 
consideration of the instantons in the $3^{4}$ boxes 
in the 4-dimensional case: it is $3^{2}=9$ boxes in the 
2-dimensional case as shown in Fig.\ref{fig:Int_M_B_2D}. 
In actual calculation, we construct link variables by 
considering all the instantons in the neighboring boxes. 

Now, we apply the MA gauge fixing \cite{Kronf1} by
maximizing 
\barr
R =  \sum_{\mu,s}\tr
[U_{\mu}(s) \tau^{3} U^{\dagger}_{\mu}(s) \tau^{3}] 
 =  2 \sum_{s,\mu} [1-2\{(U_{\mu}^{1}(s))^{2}+
(U_{\mu}^{2}(s))^{2} \}],
\label{eqn:malat}
\earr
with $U_{\mu}(s)=U_{\mu}^{0}(s)+i\tau^{i}U_{\mu}^{i}(s)$.
The maximization of $R$ corresponds to the lattice 
expression of the minimization condition of $R_{ch}[A_{\mu}]$ 
in Eq.(\ref{eqn:MAint}). In the MA gauge, the ${\rm SU(2)}$ 
link variable $U_{\mu}(s)$ is decomposed as
\barr
U_{\mu}(s)
= M_{\mu}(s)u_{\mu}(s)
= 
\left(
        \begin{array}{cc}
\sqrt{1-\mid c_{\mu}(s)\mid^{2}} &
            - c^{*}_{\mu}(s) \\
             c_{\mu}(s)          & 
\sqrt{1-\mid c_{\mu}(s)\mid^{2}}
        \end{array}
  \right)
\left(
        \begin{array}{cc}
   e^{i\theta_{\mu}(s)} &
            0           \\
            0             
  &  e^{-i\theta_{\mu}(s)}      
        \end{array}
  \right),
\earr
where the abelian angle variable $\theta_{\mu}(s)$ and 
the non-abelian variable
$c_{\mu}(s)$ are defined in terms of $U_{\mu}(s)$ as
\barr
\tan\theta_{\mu}(s) = 
\frac{~U_{\mu}^{3}(s)~}{~U_{\mu}^{0}(s)~},
\hspace{0.5cm}
 c_{\mu}(s)e^{i\theta_{\mu}(s)} 
= [-U_{\mu}^{2}(s) + i U_{\mu}^{1}(s)].
\earr
It is obvious from the expression of 
Eq.(\ref{eqn:malat}) that the off-diagonal parts 
$U_{\mu}^{1}(s)$ and $U_{\mu}^{2}(s)$ of gluon 
fields are minimized by the MA gauge transformation. 
Therefore, full ${\rm SU(2)}$ link variables would 
be approximated as ${\rm U(1)}$ link variables, 
$U_{\mu}(s) \simeq u_{\mu}(s)$, in the MA gauge. 

Monopole currents can be defined by using 
$u_{\mu}(s)$ following DeGrand and Toussaint 
\cite{DeGra}. Using a forward derivative 
$\partial_{\mu}f(s)\equiv f(s+\hat{\mu})
-f(s)$ with unit vector $\hat{\mu}$, the 2-form 
of the lattice formulation, $\theta_{\mu\nu}(s) 
\equiv \partial_{\mu}\theta_{\nu}(s)-\partial_{\nu}
\theta_{\mu}(s)$, is decomposed as
\barr
\theta_{\mu\nu}(s) = \bar{\theta}_{\mu\nu}(s) 
+ 2\pi n_{\mu\nu}(s)
\earr
with $\bar{\theta}_{\mu\nu}(s) \equiv {\rm mod}_{2\pi}
\theta_{\mu\nu} \in(-\pi, \pi]$ 
and $n_{\mu\nu}(s) \in {\bf Z}$. Here, 
$\bar{\theta}_{\mu\nu}(s)$ and $2\pi n_{\mu\nu}(s)$ 
correspond to the regular field strength and the singular 
Dirac string part, respectively. Since the abelian Bianchi 
identity is broken, the monopole current $k_{\mu}(^*\!s)$ 
can be defined on the dual link $(^*\!s,\mu)$ as
\barr
k_{\mu}(^*\!s) \equiv \frac{1}{4\pi}
\varepsilon_{\mu\nu\alpha\beta}
\partial_{\nu}\bar{\theta}_{\alpha\beta}(s+\hat{\mu}) 
=  - \partial_{\nu}\tilde{n}_{\mu\nu}(^*\!s)
\earr
where $\tilde{n}_{\mu\nu}(s) \equiv \frac{1}{2}
\varepsilon_{\mu\nu\alpha\beta}n_{\alpha\beta}
(s+\hat{\mu})$. The obvious current-conservation 
law $\partial^{\prime}_{\mu}k_{\mu}(^*\!s) = 0$
leads to the closed monopole loop in the 
4-dimensional space. Here, $\partial^{\prime}_{\mu}$ 
denotes a backward derivative. 

Since we are interested in the infrared behavior 
of monopoles in multi-instanton configurations,
it is useful to execute a ``block-spin'' 
transformation on the dual lattice with the scale 
factor ${\cal N}$. The ``block-spin'' transformation 
removes small monopole loops as numerical noises 
and keeps its global structure. Here, the 
${\cal N}^{3}$ extended monopole is defined as
\beqn
n^{\cal N}_{\alpha\beta}(s) \equiv 
\sum^{n-1}_{i,j=0} n_{\alpha\beta}(ns + i\hat{\rho} 
+ j\hat{\sigma})
\eeqn
on a sublattice with the spacing of $b={\cal N}a$ 
\cite{Ivane}. Then, the extended monopole current
is defined as $k^{\cal N}_{\mu}(^*\!s) = - \frac{1}{2}
\varepsilon_{\mu\nu\alpha\beta}\partial_{\nu} 
n^{\cal N}_{\alpha\beta}(s+\hat{\mu})$. 

We start with a very dilute instanton system
in order to clarify a local correlation between 
instantons and monopoles in the MA gauge. We take a 
density $(N/V) \ll (0.5/{\rm fm})^{4}=(100~{\rm MeV})^{4}$, 
where instantons are separated completely from each 
other. From the analytical consideration, one
may expect that a monopole loop prefers to shrink to a 
instanton center. However, we observe on the lattice that 
each finite-size monopole loop is localized around each 
instanton center as shown in Fig.\ref{fig:Ins_Mon}(a). 
Such a small monopole but finite-size loop is caused by 
finite lattice spacing and the boundary effect, which is 
numerically washed out by several numbers of ``block-spin'' 
transformation. 

In the actual QCD vacuum, instantons saturate the 
4-dimensional space and stay close to each other with a 
instanton density of $(N/V)=(1/{\rm fm})^{4}$ \cite{Shury1}. 
Therefore, we consider three instanton number-density cases, 
$(N/V)^{\frac{1}{4}}= 0.75$, $1.00$ and $1.25~{\rm fm^{-1}}$, 
and investigate the behaviors of monopole loops. 
Figure \ref{fig:His5_B}(a)
shows the histograms of unblocked monopole loops.
The histograms of blocked configurations are shown in Figs. 
\ref{fig:His5_B}(b)-(d), which corresponds to the
extended monopoles on the sublattice with $b=2a$, $b=4a$ 
and $b=8a$, respectively.

At the low instanton density where each instanton is
isolated, the monopole loop induced by the instanton 
prefers to be localized around each instanton center as 
shown in Fig.\ref{fig:Ins_Mon}(a). Thus, there appear only 
relatively short monopole loops. This situation provides a 
peak at zero monopole-loop length in the histogram as shown 
in Fig.\ref{fig:His5_B}(d-i). As the instanton density 
increases, some monopole trajectories tend to hop from 
one instanton to another nearby instanton as shown in 
Fig.\ref{fig:Ins_Mon}(b), and there appear long 
monopole-loops. Here, a clustering of long monopole 
loops appears and grows gradually to be separated 
from the small monopole-loop part in the histogram. 
Furthermore, the ``block-spin'' transformation is 
available to clarify monopole loop behaviors at the 
large scale by comparing Figs.\ref{fig:His5_B}(a-iii), 
(b-iii), (c-iii) and (d-iii). Because this procedure
removes small-size monopole loops and combine several 
long monopole-loops into one longer loop. At a high 
instanton density, there appears one very long and highly 
complicated monopole loop in each gauge configuration 
as shown in Fig.\ref{fig:His5_B}(d)-(iii). Such an
appearance of the monopole clustering over the
entire physical volume can be interpreted as the 
Kosterlitz-Thouless-type phase transition \cite{Kostr,Ichie}. 

In addition, we discuss the case of the instanton size 
distribution $\nu=3$ with the average size $\bar{\rho}= 
0.4~{\rm fm}$ as shown in Fig.\ref{fig:Size_Dis}. Diakonov 
pointed out that if the instanton size distribution falls off 
as $f(\rho)\sim 1/\rho^{3}$ in the infrared region, one 
gets a linear confinement potential \cite{Diako2},
which is based on the formula for the inter-quark potential 
in Ref. \cite{Diako7}. Since instantons tend to have large 
overlapping from each other for the $\nu=3$ case, the simple treatment 
with the sum ansatz, which is based on the assumption of 
the statistical independence, may not be applicable for 
the construction of the multi-instanton system. 
However, the infrared behavior of the instanton size 
distribution is interesting for the confinement 
properties in the long-range region. Therefore, we 
demonstrate also the $\nu=3$ case. The 
histograms of the $\nu=3$ case are qualitatively 
similar to the corresponding histograms of the 
$\nu=5$ case, although there is a quantitative 
difference between these two cases.

Lattice QCD simulations at finite temperature suggest 
that the instanton density is largely reduced as the 
temperature increases. Now, we compare the histograms 
of the monopole loop length in the multi-instanton 
system with those of the ${\rm SU(2)}$ lattice QCD with 
$16^{3} \times 4$ at different temperatures $(\beta =2.2$ 
and $2.35)$ in Fig.\ref{fig:His_La} \cite{Sugan4}. 
The monopole-loop distribution at high instanton-density 
case as shown in Fig.\ref{fig:His5_B}(d-iii) resembles 
with the result of the confinement phase $(\beta =2.2)$
in Fig.\ref{fig:His_La}(b), where many instantons and 
anti-instantons saturate with the instanton density $(N/V) 
\simeq (1/{\rm fm})^4$. On the other hand, the dilute 
instanton system as shown in Fig. \ref{fig:His5_B}(d-i)
is similar to the result in the deconfinement 
phase in Fig. \ref{fig:His_La}(b) obtained by the lattice 
QCD $(\beta =2.35)$, where long monopole loops disappear.
This resemblance seems to indicate that 
the instanton plays a relevant role on the promotion of 
monopole loops in the confinement phase. 

\section{Color Confinement in Multi-Instanton System}
In the previous section, we have discussed that a
high-density instanton system provides highly 
complicated monopole loops. We shall further work out 
the confining property in the multi-instanton system
without referring to monopole configurations in the 
MA gauge\cite{Fukus3}. For the investigation of 
confinement properties, it is reasonable to consider 
the Wilson loop with a contour $C$, which is defined as
\beqn
W[C] \equiv {\rm tr} P \exp [ i\oint_{C}A_{\mu}dx_{\mu}]
={\rm tr}{\prod_{C}U_{\mu}(s)}.
\eeqn
If the contour $C$ is a rectangle of the dimension $T$ by $R$, 
the Wilson loop is related to the static quark potential 
\cite{Rothe} as 
\beqn
V(R)=- \lim_{T \rightarrow \infty}\{\ln W(R,T) / T\}.
\eeqn 
To extract the potential without the contamination of 
excited states, the time $T$ has to be large enough compared to 
the distance $R$ between quark and anti-quark, $T \gg R$ 
\cite{Diako4}. Hence, it is desired to take a large enough 
$T$ in the calculation of $W[C]$. In the periodic 
boundary conditions, the time distance is limited as a 
half of $N_{t}\times a$, where $N_{t}$ is the number of 
lattice in the time direction and $a$ is the lattice unit. 
Therefore, we introduce the $36^4$ lattice in the 
center of a 4-dimensional instanton configurations
without the periodic condition for the sizes, positions 
and color orientations of instantons. Here, the large Wilson 
loop can be calculated as $T\simeq 36\times a$ because 
of the non-periodic boundary. A schematic view of this 
condition sliced into 2-dimensional plane is shown in 
Fig.\ref{fig:Int_W_B_2D}. 

In our actual calculation, we consider the instanton 
configuration with the entire volume of $V = (10~{\rm 
fm})^{4}$. We have to fix now the lattice-spacing $a$. 
For instance, the instanton density $(N/V)=(1/{\rm fm})^{4}$ 
corresponds the instanton number $N=N_{+}+N_{-}=10^{4}$ in 
the volume $(10~{\rm fm})^4$. Since the gluon gauge field 
is constructed from the summation of instantons and 
anti-instantons as $A_{\mu}(s) = \sum_{N_{+},N_{-}}
(A_{I}+A_{\bar{I}})$, a fine lattice-spacing is necessary 
to justify the continuity of the link variables $U(s)=
\exp[iaA_{\mu}(s)]$ in above situation. Therefore, we 
actually take a fine lattice spacing of $a = 0.05~{\rm fm}$ 
considering the continuity, $aA_{\mu}(s) \ll 1$.

Recent lattice QCD simulations provide us the average 
instanton size $\bar{\rho}$ and the instanton density 
$(N/V)$. The instanton properties are extracted from 
the smoothed QCD vacuum by using several cooling methods. 
DeGrand \etal investigate a smoothing procedure based 
on the renormalization group equation \cite{DeGra2}.
In this ${\rm SU(2)}$ lattice calculation, they 
find the average size of instantons about $\bar{\rho}=
0.2~{\rm fm}$, which is too small a value than that of 
usual instanton liquid model, at a density of about 
$(N/V)=2~{\rm fm}^{-4}$. On the other hand, 
de~Forcrand \etal use a cooling 
algorithm based on the improved action with scale invariant 
instanton solution and provide that the instanton size 
distribution is peaked around $0.43~{\rm fm}$ \cite{DeFor2}. 
These two cooling methods provide different values about 
the average instanton size. As for the instanton size 
distribution, there are much effort to extract a size 
distribution from the lattice configurations by various 
smoothing methods \cite{Mich,DeFor2,DeGra2,Negele:1998ev}.
However, there is no consensus on the size distribution 
either. In our calculation, we take the instanton density 
as $(N/V) = (1/{\rm fm})^{4}$ and the average instanton 
size $\bar{\rho}=0.4~{\rm fm}$. These parameters are used 
in the discussion of chiral symmetry breaking \cite{Diako5} 
and give us the monopole length distribution in the range of 
those obtained by QCD simulation as shown in the previous 
section \cite{Fukus2}. For simplicity, we consider the 
typical case where the instanton number is equal to the
anti-instanton one, $N_{I} = N_{\bar{I}} = N/2$. 

Using the large $36^{4}$ lattice without periodic boundary 
condition, we can make as many as $10^{4}$ measurements 
of Wilson loops for each instanton configuration. Actually, 
we estimate the expectation value of Wilson loops from 
about $5\times 10^{6}$ measurements of Wilson loops on 
500 completely independent configurations. 
Figure \ref{fig:Wil_5_200} shows the Wilson loop for the 
case of $\nu=5$. The Wilson loop seems to decay as $\langle 
W[R,T]\rangle\propto \exp[-\sigma RT]$, which indicates the 
existence of a linear confining 
potential. As shown in Fig.\ref{fig:Pot_5_200}, the static 
quark potential $V(R)$ is proportional to the inter-quark 
distance $R$ up to the intermediate region $R \simeq 
1.2~{\rm fm}~(<T)$. Here, we have checked that the static 
potential at $R \le 1.2~{\rm fm}$ does not depend on 
$T~(\gg R)$ within errors. We speculate that instantons 
would be relevant degrees of freedom for the linear 
potential between a quark and anti-quark pair in the physically 
interesting region.

The area low behavior of the Wilson loop means that
the errors of this value grow as the area $RT$ increases, 
since the Wilson loop decreases exponentially with $RT$. 
The static quark potential is not yet demonstrated at 
longer distance region as $R > 1.2~{\rm fm}$, where it
is necessary to calculate larger Wilson loops with 
$T \gg 2.0~{\rm fm}$. To this end, we need a huge number 
of independent measurements and perform the smearing 
procedure to reduce the statistical error because of the 
small value of Wilson loops $ \langle W[R,T]\rangle\propto
\exp[-\sigma RT]$. 

Finally, we discuss the relation between the strength 
of color confinement and the instanton density.
In order to extract the string tension, we measure 
the Creutz ratio 
\beqn
\chi (R,T) \equiv \frac{~~\langle W(R,T)\rangle
\langle W(R\!-\!1,T\!\!-\!1)\rangle~~}
{~~\langle W(R,T\!\!-\!1)\rangle\langle 
W(R\!-\!1,T)\rangle~~} 
\eeqn
from the ${\rm SU(2)}$ Wilson loop.
If the logarithm of the Wilson loop can be approximated 
as $- \ln \langle W(R,T)\rangle = \sigma RT + m ( R+T )
+{\rm const}$, the Creutz ratio gives the exponent of 
the string tension, $\chi = \exp (-\sigma_{lat})$.
The dimensionless lattice string tension $\sigma_{lat}$ 
provides the physical string tension $\sigma_{phys}$ 
with the lattice-spacing $a$ as $\sigma_{phys} = 
\sigma_{lat}/a^{2}$. 

We estimate the string tension around $R \simeq 
0.8~{\rm fm}$ for each instanton density, 
$(N/V)^{\frac{1}{4}}=$ $0.50$, $0.75$ and $1.0~{\rm fm}^{-1}$. 
As shown in Fig.\ref{fig:Str_dim5}, the string tension 
depends directly on the instanton density, and grows 
drastically as the instanton density increases.
At a density of $(N/V)=(1/{\rm fm})^{4}$ with $\nu=5$, 
the string tension comes out to be about $\sigma \simeq 
0.4~{\rm GeV/fm}$, which corresponds to about a half 
of the physical string tension $\sigma_{exp} \simeq 0.89 
{\rm GeV/fm}$. 

In addition, we demonstrate the $\nu=3$ case. As shown 
in Fig.\ref{fig:Wil_3_200} and Fig.\ref{fig:Pot_3_200},
the Wilson loop seems to obey the area low and there 
appear a linear potential at $R \le 1.2~{\rm fm}$, 
which are qualitatively similar to the $\nu=5$ case.
Figure \ref{fig:Str_dim3} shows that the string tension 
of the $\nu=3$ case is larger than that of the 
$\nu=5$ case at the same density, which is caused 
by the large overlapping of instantons and 
anti-instantons in the $\nu=3$ case. Rigorously speaking,  
for $\nu=3$ it may not be suitable to adopt the 
simple sum ansatz, which is based on the statistical 
independence of the instanton ensemble. Therefore, 
one needs to construct a new formulation, which is 
workable also for the highly overlapping instanton 
system as the $\nu=3$ case.

Finally, we would like to discuss the difference 
of the present results with the 
previous works. In our previous calculation \cite{Fukus3}, 
we fixed the peak of the instanton size distribution as 
$\rho_{peak}=0.4~{\rm fm}$, which corresponds to the 
average instanton size of $\bar{\rho} \simeq 0.45~{\rm 
fm}$. We extracted then the static potential using 
smaller Wilson loops. Since the string tension strongly 
depends also on the average instanton size $\bar{\rho}$, 
we obtained a larger value for the string tension 
\cite{Fukus3} than that of the present calculation. 
Recently, Brower \etal report that the slope of the heavy 
quark potential is about $0.1~{\rm GeV/fm}$ at the average 
instanton size $\bar{\rho}=1/3~{\rm fm}$ and the density 
$(N/V)=(1/{\rm fm})^{4}$ \cite{Browe2}. The main reason of 
the discrepancy is the different choice of the average 
instanton size $\bar{\rho}$. In addition, they use larger 
Wilson loops with $T$ above $3~{\rm fm}$. The large Wilson 
loop justifies the extraction of the string tension.
In this sense, they should obtain a better value for the 
string tension at long distance \cite{Browe2}. 
However, in their calculation, the perpendicular 2-dimensional 
directions $(x-y)$ to the plane $(z-t)$ of Wilson loops are 
limited to be much narrower than the size of the Wilson loop, 
although the other directions are very wide for the calculation 
of large Wilson loops. Our preliminary study of the contributions
of instantons away from the plane of the Wilson loop in the 
$x-y$ plane indicates appreciable contributions to such
a large Wilson loop. Hence, it would be necessary to consider 
a large volume even in the $x-y$ direction for the large 
Wilson loop, which is outside of the scope of this work.

\section{Summary and Conclusion}
We have studied color confinement and nonperturbative 
quantities of the QCD vacuum using the multi-instanton 
configuration. We have made the present study by being 
motivated by the presence of a strong correlation between 
instantons and monopoles after the abelian gauge fixing in 
pure ${\rm SU(2)}$ gauge theory. In our calculation, the 
multi-instanton system is constructed by assuming the 
suppression of the large size instanton as 
$f(\rho)\sim\rho^{-5}$ due to the infrared repulsive 
interaction between instantons.

First, we have investigated the monopole-loop distribution 
in the multi-instanton system by using the maximally abelian 
gauge. Here, we have executed the ``block-spin'' transformation
in order to extract the infrared properties of monopole loops.
The dilute-instanton system produces small monopole loops 
localized around each isolated instanton.
If instantons are well separated each 
other, monopole loops disappear in the continuum limit 
$a \rightarrow 0$. 
However, as the instanton density becomes higher, several 
small monopole loops combine into one longer loop. At a
high instanton density, there appears a highly complicated
monopole loop covering the entire physical volume. 
The appearance of long monopole loops of this system resembles
that of the low-temperature lattice QCD where color confinement 
is realized through monopole condensation \cite{Sugan6}. 
Our results indicate that instantons play an essential 
role on the promotion of a global network of monopoles.

We have found that the high instanton density provides a highly 
complicated and long monopole loop. This seems to indicate that 
instantons are responsible also for confinement. Therefore, 
we have calculated the Wilson loop in the multi-instanton system
without the abelian gauge fixing. 
Here, we take the instanton density of $(N/V)=(1/{\rm fm})^{4}$
and the average instanton size of $\bar{\rho}=0.4~{\rm fm}$.
We have found that 
the instanton ensemble gives an area law behavior of the Wilson 
loop and the static quark potential is approximately proportional 
to the inter-quark distance, $V(R) \simeq \sigma R$, up to 
$R \simeq 1.2~{\rm fm}$. In this situation, the string tension 
has been evaluated as $\sigma \simeq 0.4~{\rm GeV/fm}$, which 
corresponds to about a half of the physical string tension
$\sigma_{exp} \simeq 0.89 {\rm GeV/fm}$. 
Furthermore, we have discussed the dependence
of the string tension on the instanton density. The string 
tension tends to decrease monotonously as the instanton density 
becomes smaller. Such a tendency is consistent with the 
disappearance of long monopole loop as the instanton number  
decreases. At the high temperature QCD vacuum where the string 
tension is zero, instantons and anti-instantons almost disappear 
and also the long monopole loops do. From this close relation
between the monopole clustering and the instanton density, we
speculate that instantons would be relevant degrees of freedom 
for the linear potential between quark and anti-quark in the 
physically interesting region.

In the multi-instanton system, the string tension depends 
directly on both the instanton density and the average
instanton size . Therefore, one has to fix these parameters
from the original QCD vacuum. To this end, for instance, the 
improved cooling \cite{DeFor2} and the inverse-blocking \cite{DeGra2} 
are very interesting scheme for the extraction 
of topological quantities like instantons. Finally, it is 
necessary to simulate the static quark potential 
at longer distance $(R \gg 1.2~{\rm fm})$ using larger Wilson 
loops and enormous configurations in order to clarify the 
confinement properties up to very long distance. However, 
the calculation of such a large Wilson loop would be facing  
the limit of the present computation power. 

\section{Acknowledgment}
We would like to thank Dr.S.Sasaki for his useful comments and 
discussions. We would like to thank Dr.A.Tanaka for his continuous 
encouragement. One of the authors (H.S.) is supported in part by Grant 
for Scientific Research (No.09640359) from the Ministry of Education, 
Science and Culture, Japan. One of the authors (M.F.) is supported by 
Research Fellowships of the Japan Society for the Promotion of Science 
for Young Scientists. We have performed all numerical simulations 
with NEC SX4 of Osaka University.

\baselineskip 12pt

\clearpage
\begin{figure}
\begin{center}
\epsfig{figure=Ins_Siz_Dis.EPSF,height=8cm} 
\\
\vspace{1.cm}
{\Huge Fig. \ref{fig:Size_Dis}}
\end{center}
\end{figure} 

\begin{figure}
\begin{center}
\epsfig{figure=Boun_Mon_2D.EPSF,height=8cm} 
\\
\vspace{1.cm}
{\Huge Fig. \ref{fig:Int_M_B_2D}}
\end{center}
\end{figure} 
\clearpage

\begin{figure}
\begin{center}
\epsfig{figure=Ins_Mon_Low.EPSF,height=8cm} 
\\
\vspace{1.0cm}
{\Large Fig. \ref{fig:Ins_Mon}(a)}
\end{center}
\end{figure} 

\hspace{1.5cm}
\begin{figure}
\begin{center}
\epsfig{figure=Ins_Mon_Hig.EPSF,height=8cm} 
\\
\vspace{1.0cm}
{\Large Fig. \ref{fig:Ins_Mon}(b)}
\\
\vspace{1.0cm}
{\Huge Fig. \ref{fig:Ins_Mon}}
\end{center}
\end{figure} 
\clearpage

\begin{figure}[ht]
\hspace{2.5cm}
\begin{minipage}[hbt]{3.2cm}
\center{
\epsfig{figure=His5_150_B0.EPSF,height=3.2cm} 
\\
\vspace{-0.2cm}
(i)
}
\end{minipage}
\begin{minipage}[hbt]{3.2cm}
\center{
\epsfig{figure=His5_200_B0.EPSF,height=3.2cm} 
\\
\vspace{-0.2cm}
(ii)
}
\end{minipage}
\begin{minipage}[hbt]{3.2cm}
\center{
\epsfig{figure=His5_250_B0.EPSF,height=3.2cm} 
\vspace{-0.2cm}
\\
(iii)
}
\end{minipage}
\\
\vspace{-0.3cm}
\center{
{\Large Fig. \ref{fig:His5_B}(a)}}
\end{figure} 

\vspace{0.2cm}
\begin{figure}[ht]
\hspace{2.5cm}
\begin{minipage}[hbt]{3.2cm}
\center{
\epsfig{figure=His5_150_B2.EPSF,height=3.2cm} 
\\
\vspace{-0.2cm}
(i)
}
\end{minipage}
\begin{minipage}[hbt]{3.2cm}
\center{
\epsfig{figure=His5_200_B2.EPSF,height=3.2cm} 
\\
\vspace{-0.2cm}
(ii)
}
\end{minipage}
\begin{minipage}[hbt]{3.2cm}
\center{
\epsfig{figure=His5_250_B2.EPSF,height=3.2cm} 
\\
\vspace{-0.2cm}
(iii)
}
\end{minipage}
\\
\vspace{-0.3cm}
\center{
{\Large Fig. \ref{fig:His5_B}(b)}}
\end{figure} 

\vspace{0.2cm}
\begin{figure}[ht]
\hspace{2.5cm}
\begin{minipage}[hbt]{3.2cm}
\center{
\epsfig{figure=His5_150_B4.EPSF,height=3.2cm} 
\\
\vspace{-0.2cm}
(i)
}
\end{minipage}
\begin{minipage}[hbt]{3.2cm}
\center{
\epsfig{figure=His5_200_B4.EPSF,height=3.2cm} 
\\
\vspace{-0.2cm}
(ii)
}
\end{minipage}
\begin{minipage}[hbt]{3.2cm}
\center{
\epsfig{figure=His5_250_B4.EPSF,height=3.2cm} 
\\
\vspace{-0.2cm}
(iii)
}
\end{minipage}
\\
\vspace{-0.3cm}
\center{
{\Large Fig. \ref{fig:His5_B}(c)}}
\end{figure} 

\vspace{0.2cm}
\begin{figure}[ht]
\hspace{2.5cm}
\begin{minipage}[hbt]{3.2cm}
\center{
\epsfig{figure=His5_150_B8.EPSF,height=3.2cm} 
\\
\vspace{-0.2cm}
(i)
}
\end{minipage}
\begin{minipage}[hbt]{3.2cm}
\center{
\epsfig{figure=His5_200_B8.EPSF,height=3.2cm} 
\\
\vspace{-0.2cm}
(ii)
}
\end{minipage}
\begin{minipage}[hbt]{3.2cm}
\center{
\epsfig{figure=His5_250_B8.EPSF,height=3.2cm} 
\\
\vspace{-0.2cm}
(iii)
}
\end{minipage}
\\
\vspace{-0.3cm}
\center{
{\Large Fig. \ref{fig:His5_B}(d)}}
\\
\vspace{1.0cm}
{\Huge Fig. \ref{fig:His5_B}}
\end{figure}

\clearpage

\begin{figure}[ht]
\vspace{1.0cm}
\hspace{1.0cm}
\begin{minipage}[hbt]{6.0cm}
\center{
\epsfig{figure=His_Lat_Hig.EPSF,height=6.0cm}
\\
\vspace{-0.2cm}
(a)
}
\end{minipage}
\hspace{1.5cm}
\begin{minipage}[hbt]{6.0cm}
\center{
\epsfig{figure=His_Lat_Low.EPSF,height=6.0cm}
\\
\vspace{-0.2cm}
(b)
}
\end{minipage}
\center{
{\Huge Fig. \ref{fig:His_La}}}
\end{figure} 

\begin{figure}
\vspace{2.0cm}
\begin{center}
\epsfig{figure=Boun_Wil_2D.EPSF,height=8.0cm}
\\
\vspace{2.0cm}
{\Huge Fig. \ref{fig:Int_W_B_2D}}
\end{center}
\end{figure} 

\clearpage
\begin{figure}[ht]
\begin{center}
\epsfig{figure=Wils_200_D5.EPSF,height=6.5cm}
\\
{\Huge Fig. \ref{fig:Wil_5_200}}
\end{center}
\end{figure} 
\vspace{-0.5cm}
\begin{figure}[ht]
\begin{center}
\epsfig{figure=Pote_200_D5.EPSF,height=6.5cm}
\\
{\Huge Fig. \ref{fig:Pot_5_200}}
\end{center}
\end{figure} 
\vspace{-0.5cm}
\begin{figure}[ht]
\begin{center}
\epsfig{figure=Str_Tens_D5.EPSF,height=6.5cm}
\\
{\Huge Fig. \ref{fig:Str_dim5}}
\end{center}
\end{figure}

\clearpage
\begin{figure}[ht]
\begin{center}
\epsfig{figure=Wils_200_D3.EPSF,height=6.5cm}
\\
{\Huge Fig. \ref{fig:Wil_3_200}}
\end{center}
\end{figure} 
\vspace{-0.5cm}
\begin{figure}[ht]
\begin{center}
\epsfig{figure=Pote_200_D3.EPSF,height=6.5cm}
\\
{\Huge Fig. \ref{fig:Pot_3_200}}
\end{center}
\end{figure} 
\vspace{-0.5cm}
\begin{figure}[ht]
\begin{center}
\epsfig{figure=Str_Tens_D3.EPSF,height=6.5cm}
\\
{\Huge Fig. \ref{fig:Str_dim3}}
\end{center}
\end{figure}

\clearpage
\centerline{\large FIGURE CAPTIONS}

\begin{figure}
\caption{The instanton size distribution $f(\rho)$ 
as a function of the
instanton size $\rho$. The instanton size distribution for $\nu=5$
is denoted by solid curve and that for $\nu=3$ by dashed curve.
In both cases, the average size $\bar{\rho}$ is kept to $\bar{\rho}
=0.4~{\rm fm}$.}
\label{fig:Size_Dis}
\end{figure} 
\vspace{-0.3cm}

\begin{figure}
\caption{A schematic demonstration of the periodic condition 
used for an instanton ensemble 
sliced into 2-dimensional plane. We introduce the lattice in the
center of the instanton ensemble to calculate the link variables
by considering all the instantons outside of this central box
obtained with a periodic condition for the sizes, positions 
color orientations of instantons. }
\label{fig:Int_M_B_2D}
\end{figure} 
\vspace{-0.3cm}

\begin{figure}
\caption{The local correlation between instantons and monopoles 
for a dilute instanton system (a), and for a dense instanton 
system (b). The thick lines denote monopole currents which appear 
at some time $t$. The bottom surface corresponds to the action 
density of instantons $s(x,y)$ at a certain time slice. 
Here, we show monopole currents which are transformed by one 
``block-spin'' on the $32^2\times 8^2$ lattice with 
$a=0.125~{\rm fm}$.}
\label{fig:Ins_Mon}
\end{figure} 
\vspace{-0.3cm}

\begin{figure}
\caption{The histograms of monopole-loop length with various
densities $(N/V)^{1/4}=0.75, 1.0$ and $1.25~{\rm fm}^{-1}$,
after the various ``block-spin'' transformations 
in the $\nu=5$ case. (a) denote the results of no ``block-spin'' 
transformation. (b), (c) and (d) denote the results after one, 
two and three ``block-spin'' transformations, respectively.}
\label{fig:His5_B}
\end{figure} 
\vspace{-0.3cm}

\begin{figure}
\caption{The histograms of monopole-loop length
in unit of the lattice spacing $a$  
at high temperature (a), and at low temperature (b)
in finite temperature $SU(2)$ lattice QCD.}
\label{fig:His_La}
\end{figure} 
\vspace{-0.3cm}

\begin{figure}
\caption{A random instanton configuration without the 
periodic condition for sizes, positions and color 
orientations of instantons sliced into 2-dimensional 
plane. We introduce the lattice in the center of the 
random instanton ensemble. $a$ denotes the lattice unit 
and $36$ corresponds to the number of lattice point 
in each time-space direction.}
\label{fig:Int_W_B_2D}
\end{figure} 
\vspace{-0.3cm}

\begin{figure}
\caption{The ensemble average of the Wilson loop in the 
multi-instanton system with the infrared instanton size 
distribution of $\nu=5$ as a function of the area $R\times T$.
The instanton density is $(N/V)^{1/4}=1~{\rm fm}^{-1}$ and 
the average size is $ \bar{\rho} =0.4~{\rm fm}$.}
\label{fig:Wil_5_200}
\end{figure} 
\vspace{-0.3cm}

\begin{figure}
\caption{The static potential in the multi-instanton system as
a function of the distance $R$ extracted from the Wilson loop 
with $T = 1.8~{\rm fm}$ of Fig.\ref{fig:Wil_5_200}.}
\label{fig:Pot_5_200}
\end{figure} 
\vspace{-0.3cm}

\begin{figure}
\caption{The string tension $\sigma$ as
a function of the instanton density  $(N/V)^{1/4}$
with the infrared instanton size distribution $\nu=5$.}
\label{fig:Str_dim5}
\end{figure} 
\vspace{-0.3cm}

\begin{figure}
\caption{ The ensemble average of 
the Wilson loop in the multi-instanton system 
with the infrared instanton size distribution $\nu=3$
as a function of the area $R\times T$.
The instanton density and the average size are kept
same as the case of $\nu=5$.}
\label{fig:Wil_3_200}
\end{figure} 
\vspace{-0.3cm}

\begin{figure}
\caption{The static potential in the multi-instanton system as
a function of the distance $R$
extracted from the Wilson loop of Fig.\ref{fig:Wil_3_200}. }
\label{fig:Pot_3_200}
\end{figure} 
\vspace{-0.3cm}

\begin{figure}
\caption{The string tension $\sigma$ as
a function of the instanton density  $(N/V)^{1/4}$
with the infrared instanton size distribution $\nu=3$.}
\label{fig:Str_dim3}
\end{figure} 

\end{document}